\documentclass{ifacconf}

\usepackage{graphicx}      % include this line if your document contains figures
\usepackage{natbib}        % required for bibliography
\usepackage{bm}
\usepackage{amsmath}
\usepackage{amssymb}
\usepackage{color}
\usepackage{mathtools}
\usepackage{caption}
\usepackage{multirow}
\usepackage{tikz}
\usetikzlibrary{calc}

\newcommand{\real}{\mathbb{R}}
\newcommand{\nonnegativeInt}{\mathbb{N}}
\newcommand{\proba}{\mathbb{P}}
\newcommand{\dimsys}{d_{\mathrm{S}}}
\newcommand{\coeffindices}{\{0,1,\ldots,\dimsys\}}
\newcommand{\dimtru}{N_{\mathrm{T}}}
\newcommand{\coordx}{p_{\mathrm{x}}} % vehicle coordinate x
\newcommand{\coordy}{p_{\mathrm{y}}} % vehicle coordinate y
\newcommand{\dotcoordx}{\dot{p}_{\mathrm{x}}} % vehicle coordinate x
\newcommand{\dotcoordy}{\dot{p}_{\mathrm{y}}} % vehicle coordinate y
\newcommand{\eqdefn}{\triangleq} % definitions in equations:
\newcommand{\sumdim}[2]{{#1}^{{\leq}#2}}
\newcommand{\vectordim}[1]{\in \real^{#1}}
\newcommand{\matrixdim}[2]{\in \real^{{#1} \times {#2}}}
\newcommand{\E}{\mathbb{E}}
\newcommand{\OOO}{\mathcal{O}}
\newcommand{\ens}[1]{\left\{{#1}\right\}}
\newcommand{\enscomp}[2]{\left\{{#1}\,\middle\vert\,{#2}\right\}}
\newcommand{\trunc}[1]{\widetilde{#1}}
\newcommand{\abs}[1]{| {#1} |}
\newcommand{\card}[1]{| {#1} |}
\newcommand{\norm}[1]{\| {#1} \|}
\newcommand{\captionspace}{\vskip -10pt}
\DeclareMathOperator{\MC}{MC}

\renewcommand{\epsilon}{\varepsilon}
\renewcommand{\phi}{\varphi}

\newcommand{\mytablecaption}[2]{\refstepcounter{table}\label{#2}
  \raisebox{7pt}{\normalsize{Table \thetable.\ #1}}}

\newcommand{\mycaption}[2]{\refstepcounter{figure}\label{#2}\raisebox{-7pt}
  {Fig. \thefigure.\hspace{3pt} #1}}

\usepackage{subfig}

\AtBeginDocument{%
	\setlength\abovedisplayskip{6pt}
	\setlength\belowdisplayskip{6pt}}
\begin{document}
{\onecolumn\large 
\noindent  \textcopyright  2020 the authors.  %This work has been submitted to IFAC for possible publication.
This work has been accepted to IFAC for publication under a Creative Commons Licence CC-BY-NC-ND.

}
\newpage
\twocolumn
	\begin{frontmatter}
		
		\title{Moment Propagation of Discrete-Time Stochastic Polynomial Systems using Truncated Carleman Linearization}
		
		\thanks[footnoteinfo]{The authors are supported by ERATO HASUO
			Metamathematics for Systems Design Project No.~JPMJER1603, JST;
			J. Dubut is also supported by Grant-in-aid No.~19K20215, JSPS.}
		
		\author[First]{Sasinee Pruekprasert}
		\author[First]{Toru Takisaka}
		\author[First,Second]{Clovis Eberhart}
		\author[First]{Ahmet Cetinkaya}
		\author[First,Second]{J\'er\'emy Dubut}
		
		\address[First]{National Institute of Informatics, Tokyo, Japan
			\\(e-mail: \{sasinee, cetinkaya, eberhart, takisaka, dubut\}@nii.ac.jp).}
		\address[Second]{Japanese-French Laboratory of Informatics, UMI 3527, CNRS, Tokyo, Japan}
		
		\begin{abstract}                % Abstract of not more than 250 words.
			We propose a method to compute an approximation of the moments of a discrete-time stochastic polynomial system. We use the Carleman linearization technique to transform this finite-dimensional polynomial system into an infinite-dimensional linear one. After taking expectation and truncating the induced deterministic dynamics, we obtain a finite-dimensional linear deterministic system, which we then use to iteratively compute approximations of the moments of the original polynomial system at different time steps. We provide upper bounds on the approximation error for each moment and show that, for large enough truncation limits, the proposed method precisely computes moments for sufficiently small degrees and numbers of time steps. We use our proposed method for safety analysis to compute bounds on the probability of the system state being outside a given safety region. Finally, we illustrate our results on two concrete examples, a stochastic logistic map and a vehicle dynamics under stochastic disturbance.
		\end{abstract}
		
		
	\end{frontmatter}
	%===============================================================================
	
	\section{Introduction}
	
	\begin{figure*}
	\begin{center}
		\begin{tikzpicture}
		\node[draw,align=center] (1) at (0,0) {finite dim.\\polynomial\\stochastic};
		\node[draw,align=center] (2) at (5,0) {infinite dim.\\linear\\stochastic};
		\node[draw,align=center] (3) at (10,0) {infinite dim.\\linear\\deterministic};
		\node[draw,align=center] (4) at (15,0) {finite dim.\\linear\\deterministic};
		\path[->] (1) edge node[above]{Carleman lin.} node[below]{Section~\ref{sec:CarlemanLin}} (2);
		\path[->] (2) edge node[above]{Expectation} node[below]{Section~\ref{sec:MomentEq}} (3);
		\path[->] (3) edge node[above]{Truncation} node[below]{Section~\ref{sec:TruncatedSys}} (4);
		\end{tikzpicture}
		\mycaption{The different steps of the method}{fig:steps}
	\end{center}
	\end{figure*}
	
	Ensuring safety in cyber-physical systems under uncertainties is an
	important and challenging task. For instance, automated driving
	systems are required to take into account the uncertainties in the
	environment and those in control operations to avoid collision with
	obstacles~\citep{BryR11rapidly}. Motion planning algorithms in
	automated driving typically compute probabilities of collision in
	different potential paths of a vehicle by identifying possible future
	states of the vehicle under the effects of measurement and actuation
	uncertainties.
	
	To quantify how uncertainties affect the state of the system over time,
	one often uses the idea of \emph{mean and covariance propagation}.
	Particularly, if the dynamical system is linear and subject
	to additive Gaussian noise, then this state follows a Gaussian
	distribution, and the evolution of its mean and covariance can be described
	by a linear equation. Given the initial mean and covariance, one can
	then use this linear equation to compute mean and covariance iteratively
	at each future time instant. This idea has been used in conjunction with
	a Kalman-filtering approach in motion planning by \cite{BryR11rapidly}
	and \cite{BanzhafDSZ18footprints}. Here, we are interested in
	whether similar iterative techniques can be used to compute mean,
  	covariance, and even higher moments when the
	dynamical system is nonlinear and the noise is not necessarily additive
	and Gaussian. We provide a positive answer to this question.
	
	In this paper, we propose a method to compute approximations of the
  moments through truncated Carleman
  linearization~\citep{steeb1980non}.  Specifically, we consider
	a discrete-time nonlinear stochastic polynomial system with a model
	that allows us to describe additive and multiplicative stochastic
	noise in a unified manner through randomly varying coefficient matrices.
	We use the Carleman linearization technique on this system
	to obtain an infinite-dimensional linear one that characterizes
	the evolution of all Kronecker powers of the state variables. We observe
	that the expected dynamics of this linear system relates the moments
	of the random coefficient matrices to the moments of the state.
	By using a truncation approach, we then obtain a finite-dimensional
	linear system, which we use to compute an approximation of the mean and
	covariance of the system state and its higher moments up to
  a given truncation limit.
	
	We then propose a framework to verify the stochastic safety of the
  system via tail probability estimation
  techniques~\citep{gray1991general}.
	This framework relies on the efficient computation of error bounds
  for our moment approximations, which we provide.
  The exact error is the sum of many terms whose number grows quickly 
  with respect to the dimension of the system and time,
  and we give an efficient way to compute a bound by splitting
  the sum into two parts: the part (e.g., of linear or polynomial size in the dimension of the
  system) of the sum that strongly contribute to the error and
  on which we make a fine computation, and the rest on which we do a
  cruder but faster computation.
	
	%% Short version:
	%% We demonstrate the utility of our approach with two case
	%% studies. First, we compute moments for the scalar stochastic logistic
	%% map with uniformly distributed random growth/decay rates.  We then
	%% look at a higher-dimensional polynomial discrete-time vehicle model
	%% obtained via second-order Taylor expansion. By employing our
	%% approximate moment computation approach, we investigate how the
	%% uncertainty in the vehicle acceleration affects its future
	%% positions.
	
	%% Long version, with receding horizon framework
	We demonstrate the utility of our approach with two case
	studies. First, we compute moments for the scalar stochastic logistic
	map with uniformly distributed random growth/decay rates.  We then
	look at a higher-dimensional polynomial discrete-time vehicle model
	obtained via second-order Taylor expansion of a kinematic bicycle model. 
	We use our approach to investigate how the
	uncertainty in the acceleration affects the vehicle's future
	positions. In particular, our approach uses the moments of the
	initial state of the vehicle, together with the moments of its noisy
	acceleration. The moments of the initial state of the vehicle is
	obtained by using a noisy measurement of the state and the information
	on the statistical properties of the measurement error. Our framework
	allows moment computation in a receding horizon fashion. Specifically,
	at each time step, new measurements can be used for specifying the
	moments of the new ``initial'' state, which can then be used to compute the
	moments associated with time instants that are further in the future.

	Carleman linearization is a well-known approach in nonlinear dynamical
	systems literature.  It has been used in several pieces of work for 
approximation~\citep{bellman1963some,steeb1980non,tuwaim1998discretization},
	and more recently for optimal control~\citep{amini2019carleman} and model
	predictive control of continuous-time deterministic nonlinear
	systems~\citep{hashemian2019feedback}. Results on Carleman linearization for
	stochastic systems are relatively scarce. For Ito-type
	stochastic differential equations with bilinear noise,
	\cite{wong1983carleman} used Carleman linearization to obtain
	differential equations describing the evolution of the
	moments. Moreover, \cite{rauh2009carleman} used a series-expansion
	approach together with the Carleman linearization technique to approximate a
	continuous-time nonlinear stochastic system with a discrete-time
	linear system under additive noise. They then use the linear system
	for state estimation via a Kalman filter. Recently,
	\cite{cacace2014carleman,cacace2017optimal} developed Carleman
	linearization-based filters for nonlinear continuous-time
	stochastic systems with Wiener noise and periodic measurements. Although
	the discrete-time representation of the sampled-data filter equations in those
	pieces of work involves multiplicative noise, the models and the problem 
	formulations there differ from those in our paper: on the one hand, they can
	describe non-polynomial dynamics, but on the other hand, they can only
	handle Wiener noises. We note
	that our truncation error analysis is partly motivated by the analysis
	in \cite{forets2017explicit}, which provides tight approximation
	error bounds for the truncated Carleman linearization approach in deterministic
	continuous-time systems. We also note that in addition to Carleman
	linearization, Koopman operators are also used for deriving
	infinite-dimensional linear equations to describe finite-dimensional
	nonlinear systems~\cite{goswami2017global,mesbahi2019modal},
  although the techniques for analysis of these two methods are quite different.

	The rest of the paper is organized as follows
  (see also Figure~\ref{fig:steps} for the specific subsections where
  the main steps of the method are presented).
  In Section~\ref{sec:Carleman}, we start from a finite-dimensional
  stochastic polynomial system, on which we apply the Carleman
  linearization technique to get an infinite-dimensional linear
  stochastic system, which we then turn into a deterministic one by
  taking expectation.
	We present our truncation-based moment approximation method
	and study the error it introduces in Section~\ref{sec:Approximation},
  in which we also give an application of this bound.
	We provide two numerical examples to demonstrate our approach
	in Section~\ref{sec:numerical-example}.
	Finally, in Section~\ref{sec:Conclusion}, we conclude the paper.
	
	\textbf{Notations.} We use $\nonnegativeInt$ for the set of
  non-negative integers.
  We denote by $M_1\otimes M_2$ the Kronecker product of two matrices
	$M_1$ and $M_2$. Moreover, we use $M^{[k]}$ to denote the $k$th
  Kronecker power of the matrix $M$, given by $M^{[0]} = 1$ and
	$M^{[k]}=M^{[k-1]}\otimes M$ for $k > 0$.
	The dimensions of the matrices we use grow rapidly, so we will use
	$\sumdim{n}{k}$ as a shorthand for $\sum_{i = 0}^k n^i$ when
	manipulating dimensions of vectors or matrices.
	
	\section{Carleman Linearization for Stochastic Polynomial Systems}
	\label{sec:Carleman}
	
	In this section, we use the Carleman linearization technique to turn
  a finite-dimensional discrete-time stochastic polynomial system into
  an infinite-dimensional one, then take expectation to get a
  deterministic system that describes the evolution of all moments of
  the system state.
	
	\subsection{Discrete-Time Stochastic Polynomial Systems}
	
	We focus on computing the moments of the state of the
	following finite-dimensional discrete-time stochastic polynomial
	system
	\begin{align}
	\label{eq:system}
	\begin{split}
	x(t+1) &= \sum\limits_{i = 0}^{\dimsys} F_i(t)x^{[i]}(t), \quad t\in\nonnegativeInt,\\
	x(0) &=x_0,
	\end{split}
	\end{align}
	where $x(t)\in \real^n$ is the state vector and $F_i(t) \in
  \real^{n\times n^i}$, $i\in\coeffindices$, are randomly-varying
  coefficient matrices.
  To be more precise, we assume given a probabilistic space $\Omega$,
  then $x_0$ (\emph{resp.}\ $x(t)$, $F_i(t)$) is actually a measurable
  function from $\Omega$ to $\real^n$ (\emph{resp.}\ $\real^n$,
  $\real^{n\times n^i}$).
  We will omit to mention $\Omega$ in this paper.

	We remark that \eqref{eq:system} is useful for modeling dynamics with both 
	additive and multiplicative noise terms: the vector $F_0(t)\in \real^n$ 
	represents additive noise, as $x^{[0]}=1$, while the terms 
	$F_1(t),\ldots,F_{\dimsys}(t)$ characterize the effects of multiplicative noise.
	
	\subsection{Carleman Linearization}
	\label{sec:CarlemanLin}
	We use Carleman linearization to obtain an infinite-dimensional
  linear system that describes the evolution of the Kronecker powers
  of the state $x(t)$.
	By defining
	$$F(t) \eqdefn
	\begin{bmatrix}
	F_0(t) & F_1(t) & F_2(t) &\cdots & F_{\dimsys}(t)
	\end{bmatrix}
	\matrixdim{n}{\sumdim{n}{d_S}}$$
	and
	\begin{align}
	y(k,t) \eqdefn
	\begin{bmatrix}
	1 & x(t)^\intercal & x^{[2]}(t)^\intercal &
	\cdots & x^{[k]}(t)^\intercal
	\end{bmatrix}^\intercal \vectordim{\sumdim{n}{k}}, \label{eq:ydef}
	\end{align}
	the dynamical system \eqref{eq:system} can be rewritten as
	\begin{align*}
	x(t+1) = F(t)y(\dimsys,t).
	\end{align*}
	
	Therefore, for all $j \in \nonnegativeInt$, we have
	\begin{align*}
	  x^{[j]}(t+1) = (F(t)y(\dimsys,t))^{[j]}.
	\end{align*}
	
	By using the \emph{mixed-product property} of the Kronecker product
	(i.e., $(A \otimes B)(C \otimes D) = AC \otimes BD$; see Section~13.2
	of \cite{laub2005matrix}), we obtain
	\begin{align*}
	  x^{[j]}(t+1)
      &= F ^{[j]}(t)y^{[j]}(\dimsys,t) \\
      & = \sum_{k=0}^{j\dimsys} \left(
      \smashoperator[r]{\sum_{(i_l)_{l \leq j} \in H_{j,k}}}
      F_{i_1}(t) \otimes \cdots \otimes F_{i_j}(t) \right) x^{[k]}(t),
	\end{align*}
	for $j\in\nonnegativeInt$, where 
  $$H_{j,k} \eqdefn \enscomp{(i_l)_{l \leq j}}{\sum\limits_{l=1}^j i_l
  = k \text{ and } i_l\leq \dimsys}.$$
	We thus get the infinite-dimensional linear system
	\begin{align}
	\begin{split}
    x^{[j]}(t+1) &= \sum_{k=0}^{j\dimsys} \left( \sum_{(i_l)_{l \leq j} \in
		  H_{j,k}} \bigotimes_{m = 1}^j F_{i_m}(t) \right) x^{[k]}(t), \\
	x^{[j]}(0) &= x^{[j]}_0,
	\end{split}
	\label{eq:mixed-product}
	\end{align}
	which describes the evolution of all Kronecker powers $x^{[0]}(t),
	x^{[1]}(t), \ldots$ of the state $x(t)$.
	
	In order to give a simpler description of the system, we introduce the
	matrix $A_{j,k}(t) \in \real^{n^j \times n^k}$ given by
	\begin{align*}
	A_{j,k}(t) \eqdefn \sum_{(i_l)_{l\leq j}\in H_{j,k}} F_{i_{1}}(t)
	\otimes \cdots \otimes F_{i_j}(t).
	\end{align*}
	Note in particular that $A_{0,0}(t) = 1$ since there is only the empty
	sequence in $H_{0,0}$, and $A_{j,k}(t) = 0$ if $j = 0$ and $k > 0$ or
	$k > j \dimsys$ since $H_{j,k}$ is then empty.
	We also introduce, for all $N, M \in \mathbb{N}$, the matrix $A(t; N,
	M)$ defined by blocks as $A(t;N,M) = [A_{j,k}(t)]_{\substack{j \leq N
			\\ k \leq M}}$, that is,
	\begin{align*}
	A(t;N,M) & \triangleq\left[\begin{array}{ccc}
	A_{0,0}(t) & \cdots & A_{0,M}(t)\\
	\vdots & \ddots & \vdots\\
	A_{N,0}(t) & \cdots & A_{N,M}(t)
	\end{array}\right] \matrixdim{\sumdim{n}{N}}{\sumdim{n}{M}}.
	\end{align*}
	
	Then, for any $k \in \mathbb{N}$, we have
	\begin{align} \label{eq A k x k^2}
	y(k,t+1)
	&=
    \begin{bmatrix}
	1 & x(t+1)^\intercal & x^{[2]}(t+1)^\intercal &
	\cdots & x^{[k]}(t+1)^\intercal
    \end{bmatrix} \nonumber \\
	&= A(t;k,k\dimsys)
    \begin{bmatrix}
	1 & x(t)^\intercal & x^{[2]}(t)^\intercal &
	\cdots & x^{[k\dimsys]}(t)^\intercal
    \end{bmatrix} \nonumber \\
	&= A(t;k,k\dimsys)y(k\dimsys,t).
	\end{align}

	\subsection{Moment Equations}
  \label{sec:MomentEq}

	We now derive the deterministic system that describes the evolution of
	the moments of $x(t)$ by taking expectation in~\eqref{eq A k x k^2}.
	This gives
	\begin{align*}
	\mathbb{E}[y(k,t+1)] &= \mathbb{E}[A(t;k,k\dimsys)y(k\dimsys,t)].
	\end{align*}

  We make the following assumptions concerning the coefficient matrices and the random initial state $x_0$.
	\begin{assum} \label{assum:f-independence}
    The matrices $\enscomp{F(t)}{t\in\nonnegativeInt}$ are independent.
    Moreover, $x_0$ is independent of $F(t)$, $t\in\nonnegativeInt$. 
	\end{assum}
	
	\begin{assum} \label{assum:f-identical}
    The matrices $\enscomp{F(t)}{t\in\nonnegativeInt}$ are identically
    distributed.
  \end{assum}
	
	Note that Assumptions~\ref{assum:f-independence} and~\ref{assum:f-identical}
	 are not overly restrictive and they hold in fairly general situations as we 
	 discuss in Section~\ref{sec:numerical-example}. 
	 Notice also that under Assumption~\ref{assum:f-independence}, matrices 
	 $F_i(t)$ and $F_j(t)$ are still allowed to statistically depend on each other. 
	 Moreover, Assumption~\ref{assum:f-identical} allows us to obtain a 
	 ``time-invariant'' method to compute moments, further yielding computational 
	 advantage.

	By iteration of \eqref{eq A k x k^2}, 
	we get that $y(k\dimsys,t)$ is given by
	\begin{align}
	y(k\dimsys,t) & = A(t-1;k\dimsys,k\dimsys^2)\cdots A(0;k\dimsys^{t},k\dimsys^{t+1}) \nonumber \\
	& \quad \cdot y(k\dimsys^{t+1},0),\quad t\in \nonnegativeInt. \label{eq:ykdsatt}
	\end{align}

	It follows from Assumption~\ref{assum:f-independence} that 
	$A(t;k,k\dimsys)$ and $y(k\dimsys,t)$ in \eqref{eq A k x k^2} are mutually
	 independent. To see this, observe that $A(t;k,k\dimsys)$ is composed of 
	 the matrices $F_i(t)$, which are independent of $x_0$ and 
	 $F(t-1),\ldots,F(0)$, which determine $y(k\dimsys,t)$
	 as given by \eqref{eq:ykdsatt}.
	
	It then follows that
	\begin{align*}
	\mathbb{E}[y(k,t+1)] &= \mathbb{E}[A(t;k,k\dimsys)]\mathbb{E}[y(k\dimsys,t)].
	\end{align*}
	
	Here again, to give a simpler description of the system, we introduce
	new matrices.
	Notice that the coefficients of the matrix $\mathbb{E}[A(t;N,M)]$ are
	products of the moments of coefficients of $F(t)$ and thus independent
	of $t$ by Assumption~\ref{assum:f-identical}.
	We denote this matrix by $E(N,M)
	\matrixdim{\sumdim{n}{N}}{\sumdim{n}{M}}$, emphasizing the fact that
	it is independent of the time.
	Similarly, we denote by $E_{i,j} \matrixdim{n^i}{n^j}$ the matrix
	$\mathbb{E}[A_{i,j}(t)]$.
	The equation above can thus be rewritten to give the following system
	\begin{align} \label{eq:momentsequation}
	\begin{split}
	\mathbb{E}[x^{[j]}(t+1)] &= \sum\limits_{k = 0}^{j\dimsys} E_{j,k}\mathbb{E}[x^{[k]}(t)],\quad t\in \nonnegativeInt, \\
	\mathbb{E}[x^{[j]}(0)] &= \mathbb{E}[x_0^{[j]}].
	\end{split}
	\end{align}
	
	Since the matrices $E_{j,k}$ only depend on the moments of the
  matrices $F(\cdot)$, they can be computed offline.
	
	\section{Moment Approximation through Truncation}
	\label{sec:Approximation}
	
	We will now introduce an approach that allows us to compute
	\emph{approximations} of the moments of $x(t)$.
	This truncation approach is critical, as an exact computation of the
	moments is impossible from a practical point of view.
	Indeed, by~\eqref{eq:ydef}, computing the first $k$ moments
	at time $t$ amounts to computing $\mathbb{E}[y(k,t)]$. So, by
  	iteration of~\eqref{eq:momentsequation},
	$$\mathbb{E}[y(k,t)] = E(k,k\dimsys)\cdots E(k\dimsys^{t-1},k\dimsys^t)\mathbb{E}[y(k\dimsys^t,0)].$$
	As
	this indicates, exact computation of the first $k$ moments requires
	the knowledge of matrices $E(k,k\dimsys)$, $E(k\dimsys,k\dimsys^2)$,
	$\ldots$, $E(k\dimsys^{t-1},k\dimsys^t)$ of exponentially increasing
	dimensions, making any practical computation unrealistic.
	We thus resort to a truncation approach where we fix a truncation
	limit and consider a matrix of fixed size to approximately compute
  the moments of $x(t)$.
	
	\subsection{Approximate Moments and the Truncated System}
	\label{sec:TruncatedSys}
	
	In this section, we define the system that we use to compute
	approximations of the moments of $x(t)$.
	
	We fix $\dimtru \in \mathbb{N}$ and define $\widetilde{x}_i(t) \in \real^{n^i}$,
	$i\in{1,\ldots,\dimtru}$, by
	\begin{align}\label{eq tilde x}
    & \begin{bmatrix}
      1 & {\widetilde{x}_1(t)}^\intercal & % \widetilde{x}_2(t) &
      \cdots  &  {\widetilde{x}_{\dimtru}(t)}^\intercal
    \end{bmatrix}^\intercal \nonumber \\
    & \quad =
    {E(\dimtru,\dimtru)^t}
    \begin{bmatrix}
      1 &  {\mathbb{E}[{x_0}]}^\intercal & % \mathbb{E}[x_0^{[2]}] &
      \cdots  & {\mathbb{E}[x_0^{[\dimtru]}]}^\intercal
    \end{bmatrix}^\intercal.
	\end{align}
	Here, the vector $\widetilde{x}_i(t)$ represents an approximation of the
	moment $\mathbb{E}[x^{[i]}(t)]$ that is computed using only our
	knowledge of the first $N_T$ moments of $x_0$.
	
	By letting
	$\trunc{y}(t)\eqdefn[\begin{array}{ccccc}
	1 & \widetilde{x}_{1}^{\intercal}(t) & \widetilde{x}_{2}^{\mathrm{\intercal}}(t) & \cdots & \widetilde{x}_{\dimtru}^{\mathrm{\intercal}}(t)\end{array}]^{\intercal}$, 
	we obtain the so-called ``truncated system'', which is a discrete-time 
	linear time-invariant system given by
	\begin{align}
	\begin{split}
	\trunc{y}(t+1) &= E(\dimtru,\dimtru)\trunc{y}(t), \quad t\in\nonnegativeInt, \\
	\trunc{y}(0) &= [\begin{array}{cccc}
	1 & \mathbb{E}[x_0(t)]^{\intercal} & \cdots & \mathbb{E}[x_0^{[\dimtru]}(t)]^{\intercal}
	\end{array}]^{\intercal}.
	\end{split}
	\end{align}
	The truncated system allows us to iteratively compute approximations
  of the moments of $x(t)$ at consecutive time instants.
	Moreover, the approach only requires an offline computation of the
	matrix $E(\dimtru, \dimtru)$.
	
  \subsection{Approximation Error Bounds}
	\label{subsec:approx_indiv}
	
	We now investigate the approximation error introduced by truncation.
	Let $e_i(t)\in\real^{n^i}$ denote the approximation error of the $i$th
	moment, that is,
	\begin{align*}
	e_i(t) \eqdefn \mathbb{E}[x^{[i]}(t)] - \widetilde{x}_i(t).
	\end{align*}
	In what follows, we provide upper bounds to $\|e_i(t)\|$ (the error
  on the $i$th moment introduced by the approximation) for
  any norm $\|\cdot\|$ induced by an inner product.
	These bounds allow us to use various techniques to study the distribution
	of the state at future time steps.
	We illustrate this in Section~\ref{subsec:tail_prob} by using tail
	probability approximations~\citep{gray1991general} to compute an upper
	bound of the probability to be outside of a given safe region.
	
	Let $j_0 \in \{0,\ldots,\dimtru\}$. Our goal is to obtain an upper bound to  $\|e_{j_0}(t)\|$.
	First, by \eqref{eq:momentsequation}, we have
%	\todo{reduce to single line for spacing?}
	\begin{align*}
	\mathbb{E}[x^{[j_0]}(t)]
	&=\sum_{j_1=0}^{j_0\dimsys} E_{j_0,j_1}\mathbb{E}[x^{[j_1]}(t-1)]\nonumber\\
	&=\sum_{j_1=0}^{j_0\dimsys} E_{j_0,j_1}\sum_{j_2=0}^{j_1\dimsys} E_{j_1,j_2} \mathbb{E}[x^{[j_2]}(t-2)]\nonumber\\
	&=\sum_{j_1=0}^{j_0\dimsys} E_{j_0,j_1}\sum_{j_2=0}^{j_1\dimsys} E_{j_1,j_2} \cdots
	\sum_{j_t=0}^{j_{t-1}\dimsys} E_{j_{t-1},j_t}\mathbb{E}[x_0^{[j_t]}],
	\end{align*}
	
	and similarly, by \eqref{eq tilde x},
	\begin{align*}
	\widetilde{x}_{j_0}(t)
	=\sum_{j_1=0}^{\dimtru} E_{j_0,j_1}\sum_{j_2=0}^{\dimtru} E_{j_1,j_2} \cdots
	\sum_{j_t=0}^{\dimtru} E_{j_{t-1},j_t}\mathbb{E}[x_0^{[j_t]}].
	\end{align*}
	
	By observing that $E_{j,k} = 0$ if $k>j\dimsys$, we obtain from the
	two equations above,
	\begin{align}
	&e_{j_0}(t) = \mathbb{E}[x^{[j_0]}(t)]  - \widetilde{x}_{j_0}(t) \nonumber\\
	&\, = \smashoperator{\sum_{j_1=\dimtru+1}^{j_0\dimsys}} E_{j_0,j_1}
	\smashoperator{\sum_{j_2=0}^{j_1\dimsys}} E_{j_1,j_2} \cdots
	\smashoperator{\sum_{j_t=0}^{j_{t-1}\dimsys}} E_{j_{t-1},j_t}
	\mathbb{E}[x_0^{[j_t]}]\nonumber\\
	&\quad + \smashoperator{\sum_{j_1=0}^{\dimtru}} E_{j_0,j_1}
	\smashoperator{\sum_{j_2=\dimtru+1}^{j_1\dimsys}} E_{j_1,j_2} \cdots
	\smashoperator{\sum_{j_t=0}^{j_{t-1}\dimsys}} E_{j_{t-1},j_t}
	\mathbb{E}[x_0^{[j_t]}]\nonumber\\
	&\quad + \cdots\nonumber\\
	&\quad + \smashoperator{\sum_{j_1=0}^{\dimtru}} E_{j_0,j_1} \cdots
	\smashoperator{\sum_{j_{t-1}=0}^{\dimtru}} E_{j_{t-2},j_{t-1}}
	\smashoperator{\sum_{j_t=\dimtru+1}^{j_{t-1}\dimsys}} E_{j_{t-1},j_t}
	\mathbb{E}[x_0^{[j_t]}].
	\label{eq:ej0}
	\end{align}
	
	As an immediate consequence, we get the following:
	\begin{prop} \label{prop:exact-computation}
		Consider the truncated approximation of the moments of
		system~\eqref{eq:system} with truncation limit $\dimtru\in
		\nonnegativeInt$.
    If $j_0 \dimsys^t \leq \dimtru$, then $\trunc{x}_{j_0}(t) = \E
    [x^{[j_0]}(t)]$.
	\end{prop}
	\begin{pf}
    We show that, if $j_0 \dimsys^t \leq \dimtru$, then
    $\norm{e_{j_0}(t)} = 0$. To this end, it is enough to show that, for
    all lines of~\eqref{eq:ej0} and sequences $(j_1,\ldots,j_t)$ of relevant indices, at
		least one $E_{j_{k-1},j_k}$ is equal to $0$.
		Let us pick the $i$th line and any sequence as above.
		In particular, $j_{i-1}\dimsys > \dimtru$, hence $j_{i-1} > j_0
		\dimsys^{t-1} \geq j_0 \dimsys^{i-1}$, so there exists $k \in
		\ens{1,\ldots,i-1}$ such that $j_k > j_{k-1} \dimsys$, which proves
		the desired result.\qed
	\end{pf}
	This shows that, for large values of truncation limit $\dimtru$, the
	proposed method computes \emph{exact} moments $\E[x^{[j_0]}(t)]$ for
	$j_0$ and $t$ small enough.
	Note that this is due to the \emph{discrete-time} nature of the
	finite-dimensional polynomial system~\eqref{eq:system}.
	In the continuous-time case, approximation errors cannot be
	avoided in general~\citep{forets2017explicit}.
	
	However, the exact value of $e_{j_0}(t)$ is generally hard to compute.
	Indeed, since $\E[x^{[i]}(t)] = \trunc{x}_{i}(t) + e_i(t)$, if
	$e_i(t)$ could efficiently be computed, then so would
	$\E[x^{[i]}(t)]$, and there would be no need in using the truncated
	system. %\todo{The first two sentences of this paragraph can perhaps be removed if we need space.}
	In the rest of this section, we thus come up with several upper bounds
	for $\norm{e_{j_0}(t)}$ that can be efficiently computed.
	
	First, observe that $\xi = \displaystyle \max_{0 \leq j \leq j_{0}
		\dimsys^t} \lVert \mathbb{E}[x_0^{[j]}] \rVert$
	can be efficiently computed in some cases.
	An obvious situation is when the position $x_0$ is determined, in
  which case we have $\xi = \max \{1, \lVert x_0 \rVert^{j_{0}\dimsys^t} \}$.
  Another case is when $x_0$ obeys a well-known distribution whose
  moments are easy to compute, such as uniform or normal distributions.
  Nevertheless, another case is when the system satisfies $x(t) \in \mathbb R$ and $0 \leq x_0 \leq 1$, in which case 
	$\lVert \mathbb{E}[x_0^{[j]}] \rVert$ is decreasing and we have 
	$\xi = \lVert \mathbb{E}[x_0^{[0]}] \rVert = 1$.
	Using $\xi$, we can derive bounds for $e_{j_0}(t)$ by first
	rewriting~\eqref{eq:ej0} as
	\begin{align}
	e_{j_0}(t) = \sum_{j=0}^{j_0\dimsys^t} \widetilde E_j \mathbb E[x_0^{[j]}], \label{eq:ej1}
	\end{align}
	where $\widetilde E_j \matrixdim{n^{j_0}}{n^j}$ is constructed by additions and
	multiplications of $E_{i,j}$, and thus can be computed
  offline (note that $\widetilde E_j$ is dependent on $t$, but we keep
  this implicit for readability).
  From this, we can derive a global bound
  \begin{align*}
    \norm{e_{j_0}(t)} \leq \xi \sum_{j=0}^{j_0\dimsys^t} \norm{\widetilde{E}_j},
  \end{align*}
  where $\sum_{j=0}^{j_0\dimsys^t} \norm{\widetilde{E}_j}$ can be computed
  offline.

  We can further refine~\eqref{eq:ej1} to consider a single line $i \leq n^{j_0}$
	of this equation, for which we get
	$
	(e_{j_0}(t))_i = \sum_{j=0}^{j_0\dimsys^t} v_{j,i} \mathbb E[x_0^{[j]}],
	$ 
	where $v_{j, i} \matrixdim{1}{n^j}$ is the $i$th row of $\widetilde{E}_j$. 
	By repeated application of triangle and Cauchy-Schwarz inequalities,
  we obtain
	\[
	| (e_{j_0}(t))_i |
	\leq \xi \sum_{j=0}^{j_0\dimsys^t} \lVert v_{j,i} \rVert,
	\]
	where $\sum_{j=0}^{j_0\dimsys^t} \lVert  v_{j,i} \rVert$ can also be
	computed offline.
	
	This bound can, however, be crude in practice as the norm gets
	distributed over all sums and products.
	Here we show how to compute tighter bounds while maintaining a
	reasonable computational cost. 
	For any subset $J \subseteq \{0, \ldots, j_0\dimsys^t\}$, we have the
	following:
	\begin{align}
	| (e_{j_0}(t))_i | 
	& \leq \Big|\sum_{j\in J}  v_{j,i} \mathbb E[x_0^{[j]}] \Big|
	+ \xi_J \sum_{j \not \in J} \lVert  v_{j,i} \rVert,  \label{eq:err_bound}
	\end{align}
	where $\xi_J = \displaystyle \max_{j \not\in J} \lVert \mathbb{E}[x_0^{[j]}] \rVert$. 
	
	The idea is that $J$ is a set of indices where one should avoid
	distributing the norm over the sum.
	One should pick $J$ to consist of those indices where
	the distribution is too crude and makes the error bound loose.
	In order for this method to be computationally efficient, one should
	pick $J$ that is of relatively small size, e.g., $\card{J} = \OOO(t)$.
	One possible way to choose $J$ is to fix a size $k$ and return the set
	of $k$ indices such that $\norm{v_{j,i}}$ is largest for those
	indices; 
	another way is to return the set of $k$ indices such that $\lVert \mathbb{E}[x_0^{[j]}] \rVert$ is largest.
	
	\begin{exmp}
	Suppose $x_0$ is drawn from a truncated normal distribution over the interval $[0,1]$. 
	Then $\xi = 1$ and $\xi_J = \lVert \mathbb{E}[x_0^{[j_{\min}]}] \rVert$, where $j_{\min}$ is the smallest number that is not in $J$. 
	In particular, if $J$ is chosen as the set of indices such that $\lVert \mathbb{E}[x_0^{[j_{\min}]}] \rVert$ is largest among $j = 0, \ldots, j_0 \dimsys^t$, then we have $J = \{0, \ldots, |J|-1\}$ and 
	$\xi_J = \lVert \mathbb{E}[x_0^{[|J|]}] \rVert$.
	\end{exmp}
	
	\subsection{Tail Probability Approximation}\label{subsec:tail_prob}

In this section, we use the error bound introduced by the
truncated system, which is derived in the previous section, to give a
lower bound on the probability of staying inside a given safe region.

Suppose we know some approximations $(\trunc{x}_2(t))_{n i + j}$
(henceforth denoted $(\trunc{x}_2(t))_{i,j}$) of expectations
$\E[x_i(t) x_j(t)]$ for all $i,j \leq n$ (where $n$ 
is the dimension of the state space, as in~\eqref{eq:system}), with respective
error bounds $\epsilon_{i,j}$, as well as approximations
$(\trunc{x}_1(t))_i$ of $\E[x_i(t)]$ with error bounds $\epsilon_i$.
Finally, assume that we know a global error bound
$\norm{\trunc{x}_1(t) - \E[x(t)]} \leq \epsilon$.
These different bounds can be found by applying the techniques in the
previous section.

We can then bound the probability to be outside of a ``safe'' region
centered on $\trunc{x}_1(t)$ %(ideally, it should be centered on
%$\E[x(t)]$, but we can only compute an approximation) as follows.
by the following proposition.

\begin{prop}\label{prop:tail_prob_analysis}
  Let $\norm{\cdot}$ be the Euclidean norm. For $\alpha > \epsilon$,
  \begin{align*}
    \proba&(\norm{x(t) - \trunc{x}_1(t)} \geq \alpha) \\
    & \leq \frac{\sum_{i=1}^n (\trunc{x}_2(t))_{i,i} + \epsilon_{i,i}
      - \max\ens{0, (\abs{(\trunc{x}_1(t))_i} -
      \epsilon_i)^2}}{(\alpha - \epsilon)^2}.
  \end{align*}
\end{prop}
\begin{pf}Since $\alpha > \epsilon$, by Chebyshev's inequality:
  \begin{align*}
    \proba(\norm{x(t) - \trunc{x}_1(t)} \geq \alpha) 
    & \leq \proba(\norm{x(t) - \E[x(t)]} \geq \alpha - \epsilon) \\
    & \leq \frac{\E[\norm{x(t) - \E[x(t)]}^2]}{(\alpha - \epsilon)^2} \\
    & = \frac{\sum_{i=1}^n \E[x_i^2(t)] - \E[x_i(t)]^2}{(\alpha -
    \epsilon)^2},
  \end{align*}
  and we conclude using $\E[x_i^2(t)] \leq (\trunc{x}_2(t))_{i,i} +
  \epsilon_{i,i}$ and $\E[x_i(t)]^2 = \abs{\E[x_i(t)]}^2 \geq \max
  \ens{0, (\abs{(\trunc{x}_1(t))_i} - \epsilon_i)^2}$. \hfill $\square$
\end{pf}

The safety region discussed in
Proposition~\ref{prop:tail_prob_analysis} corresponds to a ball in
$\real^n$ with radius $\alpha$. We note that safety regions can be
generalized to ellipsoids by changing the norm $\|\cdot\|$ to a
weighted norm $\|\cdot \|_P$ induced by a positive-definite matrix
$P\in\real^n$. Moreover, semi-norms induced by positive semi-definite
matrices can be used for problems where the safety of only certain
entries of the state vector is explored.
A similar argument may be used to prove a general formula for any
semi-norm induced by a positive semi-definite matrix, but we omit it
as it will not be used in the examples presented in this paper.

	\section{Numerical Examples} \label{sec:numerical-example}
	
    \vskip -5pt
	In this section, we provide two numerical examples and experimental
  results to illustrate our techniques.
	
	\subsection{Stochastic Logistic Map}\label{subsec:example_slm}
	
	Consider the stochastic logistic map as studied by \cite{athreya2000},
	which is given by
	\begin{align*}
	x(t+1) &= r(t) x(t) (1-x(t)), \quad t\in \nonnegativeInt, \\
	x(0) &= x_0,
	\end{align*}
	where $x_0, r(0), r(1), \ldots$ are mutually independent random
  variables; $x_0$ takes values in $[0,1]$ and $r(t)$ all take values
  in $[0,4]$.
	The scalar $x(t) \in [0,1]$ represents the population of a species
	subject to growth rate $r(t)$.
	This system can equivalently be represented by~\eqref{eq:system} with
	$\dimsys = 2$, $F_0 (t) = 0$, $F_1 (t) = r(t)$, and $F_2 (t) = -r(t)$.

  For experiments, we chose all $r(t)$ to be uniformly distributed
  over the interval $[0.3, 0.7]$, and $x_0$ to follow a normal
  distribution of mean $0.5$ and standard deviation $0.1$ truncated to
  $[0, 1]$.

\subsubsection{Moment Approximation via Truncated System.}
We first compare our moment approximations for different truncation
limits to the true value of the moments (computed using our method
with $\dimtru = 256$).
In Figure~\ref{fig:moments_logistic_map}, we plot the first and second
moments of the truncated system with different truncation limits
$\dimtru$.
In most cases, we observe that the system with the higher truncation limit gives a better approximation. 
There is one exception in the second moment approximation at $t=6$; the system with the lower limit gives a better approximation.
This behavior seems to be similar to the one 
in~\cite{forets2017explicit}, where a truncated system with a higher
$\dimtru$ diverges to infinity more quickly, while it gives a better
approximation over a longer time interval.
(In this case, this is due to the fact that the actual system is
converging to $0$, and a system with truncation limit $\dimtru$
diverges to infinity as a polynomial of degree $\dimtru$.)
We also observe that larger $\dimtru$ is required to obtain good
approximations of higher moments.
This is a natural consequence, as truncation discards more information on the dynamics of higher moments. 
		
	\subsubsection{Error Bound on Moment Approximations.}
\begin{figure}[!tb]
		\centering
			\includegraphics[width=6.8cm,clip]{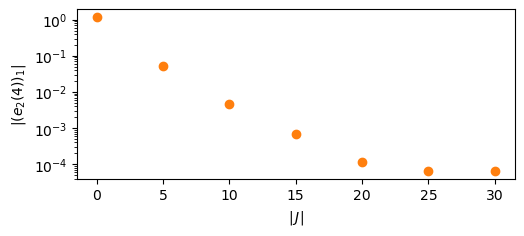}
    \captionspace
    \caption{Error bound on moment approximations.}
    \label{fig:mockup2a}
  \end{figure}

	Next, we evaluate our approximation method of error bounds for moments.
	Figure~\ref{fig:mockup2a} shows error bounds given by the
  inequality~(\ref{eq:err_bound}) with different sizes of $J$, with parameters $\dimtru = 16$, $t = 4$, and $j_0 = 2$.
	The index set $J$ contains the indices $j$ where $\lVert \mathbb{E}
  [x_0^{[j]}] \rVert$ is the largest, i.e., $J = \{0,\ldots, |J|-1\}$.
	We observe that the error bound quickly decreases as $|J|$ increases.
  This supports our expectation that we can use our error bounds for
  tail probability analysis with larger parameters ($t$, $\dimtru$,
  and $\dimsys$).
	
	\subsubsection{Tail Probability Analysis.}
\begin{figure}[!tb]
		\centering
			\includegraphics[width=7.7cm,clip]{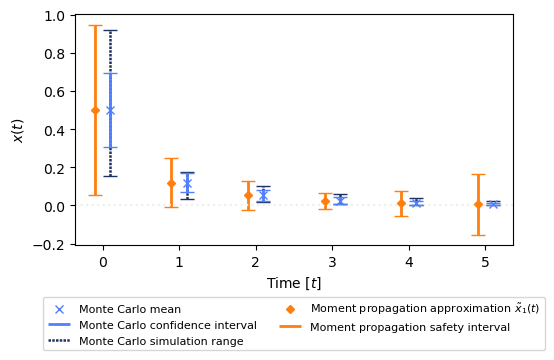}
        \captionspace
		\caption{Tail probability analysis.}
		\label{fig:mockup2b}
	\end{figure}

	Lastly, we provide a
	result on tail probability analysis via the method in Section~\ref{subsec:tail_prob}. 
	We computed the error bound for $0 \leq t \leq 5$ using~\eqref{eq:err_bound} with $\dimtru
  = 16$ and $|J| = 6 t$, where $J$ is taken in the same way as the 
  previous section.
	The results of the analysis are depicted in Figure~\ref{fig:mockup2b}.
	Orange intervals indicate the $95\%$-probability neighborhoods of
  $\trunc{x}_1(t)$ computed by using
  Proposition~\ref{prop:tail_prob_analysis}.
	Blue intervals with a solid line indicate the region where $95\%$ of
  10000 Monte Carlo simulations closest to its mean belong.
	Dotted intervals indicate the range of 10000 Monte Carlo simulations.
	
\begin{figure}[!tb]
	\centering
		\includegraphics[width=7.7cm,clip]{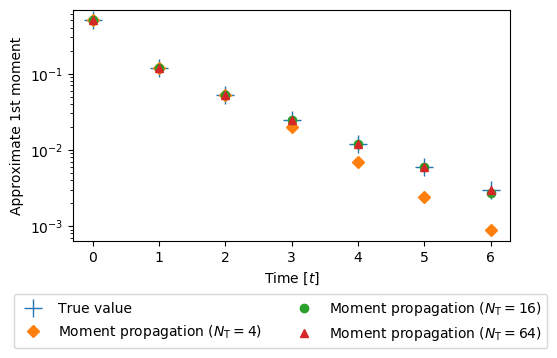}
	
		\includegraphics[width=7.7cm,clip]{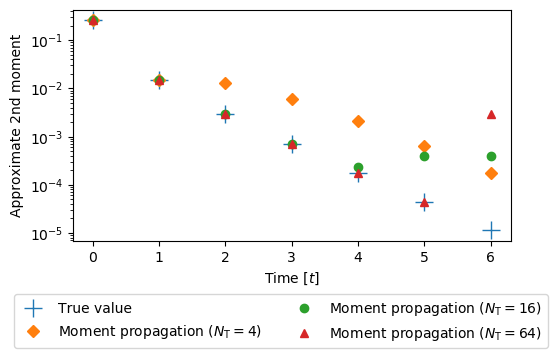}

    \captionspace
    \caption{Moment approximations for stochastic logistic map.}
	\label{fig:moments_logistic_map}
\end{figure}

	We observe that the size of safety intervals given by our tail probability analysis is reasonably small. 
	It becomes cruder in later time steps, especially at $t=5$. 
	This is expected, as the approximation error of moments, which is a bottleneck in refining the error bounds, becomes larger as time progresses (cf.\ approximate 2nd moment in Fig.~\ref{fig:moments_logistic_map}).
	
	There are two major advantages of our method compared to Monte Carlo simulation. 
	One is that our technique computes moment approximations much faster than Monte Carlo (even for large $\dimtru$ 
	and a small number of samples) because we do not rely on generating 
	random numbers.
	This advantage is highlighted in Table~\ref{table:time}, which
  contains the online computation times for Monte Carlo simulations
  and our approach, averaged over 100 runs.
	Another advantage is that our safety interval gives a theoretical guarantee on probabilistic safety that cannot be achieved by Monte Carlo simulations. 

  \begin{table}
		\mytablecaption{Comparison of online computation times.}{table:time}
    \normalsize{
    \begin{center}
      \begin{tabular}{|c|c|c|c|c|c|c|}
        \hline
        Method & \multicolumn{2}{c|}{Monte Carlo} & \multicolumn{4}{c|}{Moment propagation} \\
        \hline
        \multirow{2}{*}{Parameters} & \multicolumn{2}{c|}{num.\ samples} &
          \multicolumn{4}{c|}{$\dimtru$} \\
          \cline{2-7}
          & $10$ & % $10^2$ & $10^3$ &
          $10^4$ & $4$ & $16$ & $64$ & $256$ \\
        \hline
        Time ($\mu$s) & $2.9$e$10^3$ &
          % 2862 & % 28825 & 304320 &
          % 3358547 &
          $3.4$e$10^6$ &
          11  & 14 & 30 & 93 \\
        \hline
      \end{tabular}
    \end{center}
    }
  \end{table}

%%Figure small version
%\begin{figure}[!tb]
%	\centering
%	\subfloat[%$\mathbb{E}(x)$
%		First moment\label{fig:mockup1a}]{
%		\includegraphics[width=4.2cm,clip]{plot_approximate_first_moments_logistic_map.png}}
%	\subfloat[%$\mathbb{E}(x^2)$
%		Second moment\label{fig:mockup1b}]{
%		\includegraphics[width=4.2cm,clip]{plot_approximate_second_moments_logistic_map.png}}
%
%	\caption{Moment approximation in stochastic logistic map.}
%	\label{fig:moments_logistic_map}
%\end{figure}

	\subsection{Vehicle Dynamics}\label{subsec:example_vd}
	
  Our second example is an application to autonomous driving.  For
  safety guarantees, autonomous vehicles need to predict their
  future positions.  One way to achieve this is set-based
  reachability, as advocated by \cite{althoff2014online}.
  To use their method,
  they must consider systems with bounded disturbances and use
  linearization around an equilibrium, that is, they approximate a polynomial 
  system by a linear one, using Lagrange remainders.
  Carleman linearization allows taking the effect of higher dimensions
  of the system into account more precisely than Lagrange remainders.
  Moreover, our approach is probabilistic, while theirs is set-based,
  so the two approaches give different types of guarantees.
	
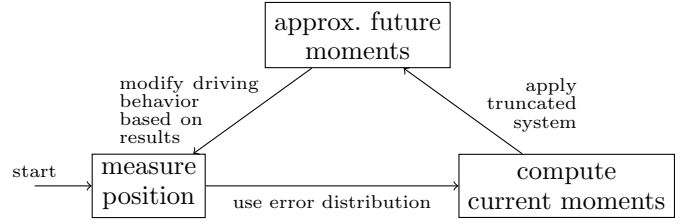
\begin{figure}[!tb]
	\begin{center}
		\begin{tikzpicture}
		\node[draw,align=center] (l) at (0,0) {measure\\position};
		\node[draw,align=center] (r) at (5.5,0) {compute\\current moments};
		\node[draw,align=center] (t) at (2.75,2) {approx.\ future\\moments};
		\node[anchor=south] () at (-1.5,0) {\scriptsize start};
		\path[->] (-1.5,0) edge (l)
              (l) edge node[below] {\scriptsize use error distribution} (r)
              (r) edge node[text width=3cm,align=right] {\scriptsize apply\\[-2pt]truncated\\[-2pt]system\\[-5pt]~} (t)
              (t) edge node[text width=3.5cm] {\scriptsize modify driving\\[-2pt]behavior\\[-2pt]based on\\[-2pt]results\\[-10pt]~} (l);
		\end{tikzpicture}
	\end{center}
    \caption{Moment propagation framework for autonomous vehicles.}
	\label{fig:autonomous_vehicle_framework}
    \end{figure}

	We consider a scenario in which, at each time step, the autonomous
	vehicle measures its current position with some known sensor error
  distributions, computes moments of its current state based on this
  measurement and the known error distribution, and predicts its
  future positions (and moments) up to $t$ steps ahead in time by
  applying the truncated system to these moments.
	This process is repeated forever, leading to the framework in Figure~\ref{fig:autonomous_vehicle_framework}.
	
	More precisely, we consider the \emph{kinematic bicycle model} of a
	vehicle~\citep{kong2015kinematic}, which we rewrite as the following
	equivalent polynomial system
	\begin{gather*}
	\dotcoordx(t) = v(t)c(t), \quad
	\dotcoordy(t) = v(t)s(t), \\
	\dot \psi(t) = \frac{v(t)}{\ell}\sin\beta, \quad
  \dot v(t) = a(t), \\
	\dot c(t) = -\frac{s(t)v(t)\sin\beta}{\ell}, \quad
	\dot s(t) = \frac{c(t)v(t)\sin\beta}{\ell},
	\end{gather*}
	for $t \geq 0$, where $\coordx (t)\in \mathbb{R}$ and $\coordy (t)\in\mathbb{R}$
	represent the X--Y coordinates of the mass-center of the vehicle,
	$v(t)\in\mathbb{R}$ denotes its speed, $\psi(t)$ its inertial
  heading, and $a(t)\in\mathbb{R}$ its acceleration.
	The constants $\beta\in\mathbb{R}$ and $\ell>0$ respectively denote
	the angle of velocity and the distance from the vehicle's rear axle to
	its mass-center.
	The scalars $c(t)$ and $s(t)$ are auxiliary variables that are
	introduced to obtain the polynomial model above from the original
	model of \cite{kong2015kinematic} (which involves
	trigonometric terms), using the same techniques
	as \cite{carothers2005some}.
	
  The second-order Taylor expansion of the model above gives the
  following discrete-time approximation:
	\begin{align*}
	\coordx(t+\Delta) &= \coordx(t) + \Delta c(t)v(t) \nonumber \\
	&\quad+ \frac{\Delta^2}{2}\biggl(a(t)c(t) - \frac{s(t)v^2(t)\sin\beta}{\ell}\biggr), \\
	\coordy(t+\Delta) &= \coordy(t) + \Delta s(t)v(t) \nonumber \\
	&\quad+ \frac{\Delta^2}{2}\biggl(a(t)s(t) + \frac{c(t)v^2(t)\sin\beta}{\ell}\biggr), \\
	\psi(t+\Delta) &= \psi(t) + \Delta\frac{v(t)}{\ell}\sin\beta + \frac{\Delta^2}{2}\frac{a(t)}{\ell}\sin\beta,\\
	v(t+\Delta) &= v(t) + \Delta a(t), \\
	c(t+\Delta) &= c(t) - \Delta\frac{s(t)v(t)\sin\beta}{\ell} \nonumber \\
	&\quad- \frac{\Delta^2}{2}\biggl(\frac{c(t)v^2(t)\sin^2\beta}{\ell^2} + \frac{a(t)s(t)\sin\beta}{\ell}\biggr),\\
	s(t+\Delta) &= s(t) + \Delta\frac{c(t)v(t)\sin\beta}{\ell} \nonumber \\
	&\quad+ \frac{\Delta^2}{2}\biggl(-\frac{s(t)v^2(t)\sin^2\beta}{\ell^2} + \frac{a(t)c(t)\sin\beta}{\ell}\biggr),
	\end{align*}
	where $\Delta>0$. To describe the evolution of the states of the vehicle at times $0, \Delta, 2\Delta, \ldots$, we write this system in the form of \eqref{eq:system}. In particular, consider the discrete-time instant $t \in \nonnegativeInt$ corresponding to the continuous time $t\Delta$. By letting
	\begin{align*}
	x(t) &\eqdefn [\coordx(t), \coordy(t), \psi(t), v(t), c(t), s(t)]^{\intercal},
	\end{align*}
	we obtain \eqref{eq:system} where $\dimsys = 3$ and the coefficients
  $F_0(t),\ldots,F_3(t)$ depend on $\Delta, \beta, \ell$, and $a(t)$.
  We consider the setting where the acceleration values $a(0), a(1), \ldots$ are
  independent uniform random variables over $[0.9,1]$, $\Delta =
  0.1$, $\beta = \pi / 8$, $\ell = 2.5$, and, for the initial state, $\coordx(0)$,
  $\coordy(0)$, $v(0)$, $\psi(0)$ are independent Gaussian random
  variables with mean $0$ and standard deviation $0.1$, and $c(0) =
  \cos(\psi(0) + \beta)$ and $s(0) = \sin(\psi(0) + \beta)$.

  \subsubsection{Experimental Results.}

	\begin{figure}[!tb]
		\centering
			\includegraphics[width=8.4cm,clip]{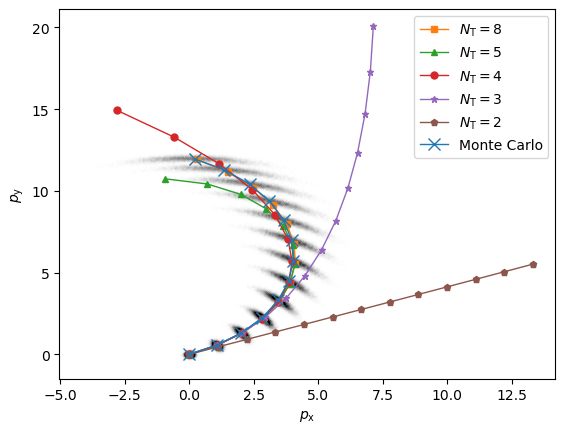}
        \captionspace
		\caption{First moment approximation of vehicle dynamics.}
		\label{fig:vehicle}
	\end{figure}
	\begin{figure}[!tb]
		\centering
			\includegraphics[width=8.4cm,clip]{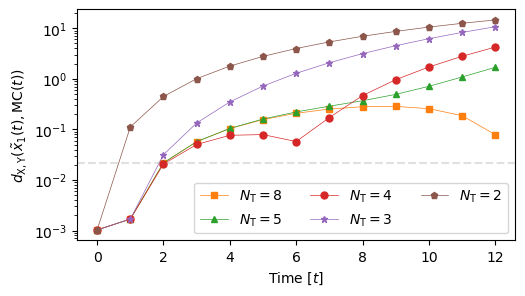}
    \captionspace
    \caption{Distance to the mean of the empirical distribution.}
		\label{fig:vehicle_distance}
	\end{figure}
  Figure~\ref{fig:vehicle} shows the expected trajectory of the
  vehicle as approximated by our method for different truncation
  limits, as well as the empirical distribution computed by 10000 runs
  of Monte Carlo simulation and the mean of that
  distribution.
  Figure~\ref{fig:vehicle_distance} shows the distance
  $d(\trunc{x}_1(t),\MC(t))$ between $\trunc{x}_1(t)$ and the mean of
  the empirical distribution for the same truncation limits.

  Figures~\ref{fig:vehicle} and~\ref{fig:vehicle_distance} show that
  larger truncation limits give systems that follow the empirical
  distribution better.
  It also shows that, for a fixed truncation limit, the distance to
  the empirical system grows larger with time.
  The only exception is around $t = 6$ for $\dimtru = 2$, which can be
  explained by the fact that the trajectory of the truncated system
  crosses that of the empirical one.
  Notice that, $\dimtru = 8$ gives an exact computation at $t = 2$,
  which means that the error at this point (corresponding to the
  horizontal line in Figure~\ref{fig:vehicle_distance}) is entirely
  induced by Monte Carlo.
  Therefore, it is pointless to consider errors that
  are of the same order as the one introduced by Monte Carlo at $t =
  2$.

  Also note that the largest truncation limit we use in this
  experiment is $\dimtru = 8$, which makes the online computation very
  fast.

	\section{Conclusion}
	\label{sec:Conclusion}
    \vskip -5pt
    We have explored the problem of computing moments of discrete-time 
	polynomial stochastic systems. Through a truncated Carleman linearization 
	approach, we have obtained an iterative procedure for approximating the 
	moments of the state of the system across different time instants. 
	Furthermore, we have provided tail-probability bounds to assess the safety 
	properties of the system.

	In our analysis, we have used spherical safety regions
  characterized using the Euclidean norm. By changing the norm to weighted
  norms induced by positive-definite matrices, ellipsoidal safety
  regions can also be considered. We note that safety analysis for a
  non-ellipsoidal region can be conducted by using ellipsoids that
  are inside that region. Finding the largest such ellipsoid allows
  us to do safety analysis in the least conservative way. This is
  possible by solving a matrix optimization problem, which we leave
  for future work.
	
	One of our future research goals is to use the approximate moments to
	reconstruct an approximation of the distribution of the state of the
	system.
	There exist different techniques to achieve this, as given for
	example by \cite{schmudgen2017moment} or \cite{john2007techniques}.
	
	Another direction for future work is to make the algorithm more
	efficient by using the symmetry of Kronecker powers
	to reduce the sizes of the matrices involved in the computations,
	which would allow us to compute even higher moments and more time
	steps in the future.

	%\bibliography{ref}             % bib file to produce the bibliography

\begin{thebibliography}{22}
\providecommand{\natexlab}[1]{#1}
\providecommand{\url}[1]{\texttt{#1}}
\providecommand{\urlprefix}{URL }
\expandafter\ifx\csname urlstyle\endcsname\relax
  \providecommand{\doi}[1]{doi:\discretionary{}{}{}#1}\else
  \providecommand{\doi}{doi:\discretionary{}{}{}\begingroup
  \urlstyle{rm}\Url}\fi

\bibitem[{{Al-Tuwaim} et~al.(1998){Al-Tuwaim}, Crisalle, and
  Svoronos}]{tuwaim1998discretization}
{Al-Tuwaim}, M.S., Crisalle, O.D., and Svoronos, S.A. (1998).
\newblock Discretization of nonlinear models using a modified {C}arleman
  linearization technique.
\newblock In \emph{Proc. Amer. Contr. Conf.}, 3084--3088.

\bibitem[{Althoff and Dolan(2014)}]{althoff2014online}
Althoff, M. and Dolan, J.M. (2014).
\newblock Online verification of automated road vehicles using reachability
  analysis.
\newblock \emph{IEEE Trans. Robotics}, 30(4), 903--918.

\bibitem[{Amini et~al.(2019)Amini, Sun, and Motee}]{amini2019carleman}
Amini, A., Sun, Q., and Motee, N. (2019).
\newblock Carleman state feedback control design of a class of nonlinear
  control systems.
\newblock In \emph{Proc. IFAC NecSys}, 229--234.

\bibitem[{Athreya and Dai(2000)}]{athreya2000}
Athreya, K.B. and Dai, J. (2000).
\newblock Random logistic maps.~{I}.
\newblock \emph{J. Theor. Probab.}, 13(2), 595--608.

\bibitem[{Banzhaf et~al.(2018)Banzhaf, Dolgov, Stellet, and
  Z{\"{o}}llner}]{BanzhafDSZ18footprints}
Banzhaf, H., Dolgov, M., Stellet, J.E., and Z{\"{o}}llner, J.M. (2018).
\newblock From footprints to beliefprints: Motion planning under uncertainty
  for maneuvering automated vehicles in dense scenarios.
\newblock In \emph{Proc. Int. Conf. Intel. Transport. Systems}, 1680--1687.

\bibitem[{Bellman and Richardson(1963)}]{bellman1963some}
Bellman, R. and Richardson, J.M. (1963).
\newblock On some questions arising in the approximate solution of nonlinear
  differential equations.
\newblock \emph{Quart. Appl. Math.}, 20(4), 333--339.

\bibitem[{Bry and Roy(2011)}]{BryR11rapidly}
Bry, A. and Roy, N. (2011).
\newblock Rapidly-exploring random belief trees for motion planning under
  uncertainty.
\newblock In \emph{Proc. IEEE {ICRA}}, 723--730.

\bibitem[{Cacace et~al.(2014)Cacace, Cusimano, Germani, and
  Palumbo}]{cacace2014carleman}
Cacace, F., Cusimano, V., Germani, A., and Palumbo, P. (2014).
\newblock A {C}arleman discretization approach to filter nonlinear stochastic
  systems with sampled measurements.
\newblock In \emph{Proc. IFAC World Congr.}, 9534--9539.

\bibitem[{Cacace et~al.(2017)Cacace, Cusimano, Germani, Palumbo, and
  Papi}]{cacace2017optimal}
Cacace, F., Cusimano, V., Germani, A., Palumbo, P., and Papi, M. (2017).
\newblock Optimal linear filter for a class of nonlinear stochastic
  differential systems with discrete measurements.
\newblock In \emph{Proc. IEEE Conf. Dec. Control}, 2807--2812.

\bibitem[{Carothers et~al.(2005)Carothers, Parker, Sochacki, and
  Warne}]{carothers2005some}
Carothers, D.C., Parker, G.E., Sochacki, J.S., and Warne, P.G. (2005).
\newblock Some properties of solutions to polynomial systems of differential
  equations.
\newblock \emph{Electr. J. Differ. Eq.}, 2005(40), 1--17.

\bibitem[{Forets and Pouly(2017)}]{forets2017explicit}
Forets, M. and Pouly, A. (2017).
\newblock Explicit error bounds for {C}arleman linearization.
\newblock \emph{Online, \url{https://arxiv.org/abs/1711.02552}}.

\bibitem[{Goswami and Paley(2017)}]{goswami2017global}
Goswami, D. and Paley, D.A. (2017).
\newblock Global bilinearization and controllability of control-affine
  nonlinear systems: {A} {K}oopman spectral approach.
\newblock In \emph{Proc. IEEE Conf. Dec. Contr.}, 6107--6112.

\bibitem[{Gray and Wang(1991)}]{gray1991general}
Gray, H.L. and Wang, S. (1991).
\newblock A general method for approximating tail probabilities.
\newblock \emph{J. Am. Stat. Assoc.}, 86(413), 159--166.

\bibitem[{Hashemian and Armaou(2019)}]{hashemian2019feedback}
Hashemian, N. and Armaou, A. (2019).
\newblock Feedback control design using model predictive control formulation
  and {C}arleman approximation method.
\newblock \emph{{AIChE} J.}, 65, 1--11.

\bibitem[{John et~al.(2007)John, Angelov, {\"O}nc{\"u}l, and
  Th{\'e}venin}]{john2007techniques}
John, V., Angelov, I., {\"O}nc{\"u}l, A., and Th{\'e}venin, D. (2007).
\newblock Techniques for the reconstruction of a distribution from a finite
  number of its moments.
\newblock \emph{Chem. Eng. Sci.}, 62(11), 2890--2904.

\bibitem[{Kong et~al.(2015)Kong, Pfeiffer, Schildbach, and
  Borrelli}]{kong2015kinematic}
Kong, J., Pfeiffer, M., Schildbach, G., and Borrelli, F. (2015).
\newblock Kinematic and dynamic vehicle models for autonomous driving control
  design.
\newblock In \emph{Proc. IEEE Intell. Veh. Symp.}, 1094--1099.

\bibitem[{Laub(2005)}]{laub2005matrix}
Laub, A.J. (2005).
\newblock \emph{Matrix Analysis for Scientists and Engineers}.
\newblock SIAM.

\bibitem[{Mesbahi et~al.(2019)Mesbahi, Bu, and Mesbahi}]{mesbahi2019modal}
Mesbahi, A., Bu, J., and Mesbahi, M. (2019).
\newblock On modal properties of the {K}oopman operator for nonlinear systems
  with symmetry.
\newblock In \emph{Proc. Amer. Contr. Conf.}, 1918--1923.

\bibitem[{Rauh et~al.(2009)Rauh, Minisini, and Aschemann}]{rauh2009carleman}
Rauh, A., Minisini, J., and Aschemann, H. (2009).
\newblock {C}arleman linearization for control and for state and disturbance
  estimation of nonlinear dynamical processes.
\newblock In \emph{Proc. IFAC Conf. MMAR}, 455--460.

\bibitem[{Schm{\"u}dgen(2017)}]{schmudgen2017moment}
Schm{\"u}dgen, K. (2017).
\newblock \emph{The Moment Problem}.
\newblock Springer.

\bibitem[{Steeb and Wilhelm(1980)}]{steeb1980non}
Steeb, W.H. and Wilhelm, F. (1980).
\newblock Non-linear autonomous systems of differential equations and
  {C}arleman linearization procedure.
\newblock \emph{J. Math. Anal. Appl.}, 77(2), 601--611.

\bibitem[{Wong(1983)}]{wong1983carleman}
Wong, W.S. (1983).
\newblock {C}arleman linearization and moment equations of nonlinear stochastic
  equations.
\newblock \emph{Stochastics}, 9(1-2), 77--101.

\end{thebibliography}
	% with bibtex (preferred)

\end{document}